\documentclass[11pt]{iopart}

\pdfoutput=1 

\usepackage{graphicx}
\usepackage{hyperref} 
\usepackage{bbold}

\newcommand{\kb}{k_\mathrm{B}}
\newcommand{\difft}{\frac{\rmd}{\rmd t}}
\newcommand{\opind}[2]{#1_\mathrm{#2}}
\newcommand{\upperind}[2]{{#1}^\mathrm{(#2)}}
\newcommand{\BSig}{\mathbf{\Sigma}}
\newcommand{\llangle}{\langle\hspace{-2.4pt}\langle}
\newcommand{\rrangle}{\rangle\hspace{-2.4pt}\rangle}

\renewcommand{\Re}{\mathop{\mathrm{Re}}}

\begin{document}

\title[Quantum heat transport in harmonic chains]{Currents and fluctuations of quantum heat transport in harmonic chains}

\author{T. Motz, J. Ankerhold and J. T. Stockburger}
\address{Institute  for Complex Quantum Systems, Ulm University - Albert-Einstein-Allee 11, D-89069  Ulm, Germany}
\ead{\mailto{thomas.motz@uni-ulm.de}, \mailto{joachim.ankerhold@uni-ulm.de}, \mailto{juergen.stockburger@uni-ulm.de}}

\begin{abstract}
Heat transport in open quantum systems is particularly susceptible to the modeling of system-reservoir interactions. It thus requires to consistently treat the coupling between a quantum system and its environment. While perturbative approaches are successfully used in fields like quantum optics and quantum information, they reveal deficiencies, typically in the context of thermodynamics, when it is essential to respect additional criteria such as fluctuation-dissipation theorems. We use a non-perturbative approach for quantum dissipative dynamics based on a stochastic Liouville-von Neumann equation to provide a very general and extremely efficient formalism for heat currents and its correlations in open harmonic chains.  Specific results are derived not only for first but also for second order moments which requires to account for both real and imaginary parts of bath-bath correlation functions. Spatiotemporal patterns are compared with weak coupling calculations. The regime of stronger system-reservoir couplings gives rise to an intimate interplay between reservoir fluctuations and heat transfer far from equilibrium.
\end{abstract}




\section{Introduction}

In the context of heat transfer, harmonic systems away from equilibrium have attracted a lot of attention over the last years since the path from first principle microscopic models to phenomenological results such as Fourier's law has turned out to be a formidable task \cite{Casher1971, OConnor1974, Bafaluy1988, Bafaluy1988a, Dhar2001, Lepri2003, Chaudhuri2010, Asadian2013}. Specifically challenging for quantum systems is the correct description of interactions between the system and environmental degrees of freedom. While in many situations the assumption of weak system-bath interactions leads to powerful perturbative techniques such as master equations, the problem of quantum heat transfer seems to be particularly susceptible to the modeling of the correlations between the relevant system and environmental degrees of freedom. In fact, the underlying assumption of an unperturbed Gibbs state of the embedded system is only valid for vanishing coupling strengths, thus implying vanishing heat transfer. Weak system-bath interactions  at low temperatures induce correlations between system and reservoir which lead to non-negligible energy exchange \cite{Ankerhold2014} and produce unphysical results in conventional perturbative treatments \cite{Levy2014, Stockburger2016}. Moreover, the relaxation towards stationary states at low temperatures generally depends on the nature of the surrounding heat baths and, particularly, on non-Markovian dynamics \cite{Grabert2006, Pagel2013, Martinez2013, Nicacio2015}. Since harmonic systems allow, at least in principle, for exact results, they may on the one hand serve as non-trivial paradigmatic examples and on the other hand as starting points for more elaborate models.

Quantum heat transport through harmonic chains has been treated based on quantum Langevin equations which for linear systems leads
 to correct results for a large variety of couplings and temperatures \cite{Zurcher1990, Zurcher1990b, Gaul2007}. However, while conceptually simple, practically, these approaches focus on symmetrized bath correlation functions to avoid the sampling of quantum time evolutions in presence of complex-valued quantum noise. They are thus restricted to study symmetrized correlation functions, which reflect the dispersive part of the system and are connected with the fluctuation-dissipation theorem to the anti-symmetric correlation that constitutes the absorptive part \cite{Callen1951}. These symmetrized correlations are sufficient to extract first moments of heat fluxes, but do not allow to gain access to higher order moments which are determined also by contributions from phase-space correlations that are not symmetrized \cite{Saito2007,PagelThor2013}. Further,  these approaches represent the dynamics in terms of normal modes of a linear chain while the system-bath coupling remains local. Their numerical efficiency thus degrades substantially for sufficiently long chains.
The aim of this paper is to provide a complete formalism which combines a non-perturbative treatment of the system-bath interaction with a high computational efficiency and allows to study chains exposed to any damping, reservoir temperatures, and in presence of external driving.

It is based on a description of open quantum dynamics in terms of reduced density operators as pioneered by Feynman and Vernon. The corresponding formally exact path integral expression allows for an exact mapping on  a numerically more convenient stochastic Liouville-von Neumann equation (SLN) \cite{Stockburger1998, Stockburger1999}. The SLN is particularly beneficial to include external time dependent forces and applies to any coupling strength and bath temperatures. Technically, system-bath interactions are exactly accounted for by introducing stochastic $c$-noises \cite{Schmidt2011} the correlations of which reproduce the exact quantum bath-bath correlations. Here, we start from this very general formalism and adapt it to describe heat transfer through harmonic chains. Accordingly, the stochastic sampling can explicitly be performed to arrive at a deterministic equation of motion for the covariance matrix. The corresponding set of coupled ordinary differential equations can be solved in a straightforward manner even for a very large number of constituents in the chain.

Specifically, we apply the formalism to harmonic chains as shown in figure~\ref{fig:osci_chain} where the system is out of equilibrium due to the coupling on reservoirs at different temperatures. The transient time evolution as well as nonequilibrium steady states for heat fluxes and their correlations for a large variety of dampings, couplings and temperatures is investigated. A high sensitivity of  heat fluxes on damping strengths and intra-chain couplings is observed. Further insight is obtained by considering spatiotemporal heat-flux correlations which are compared with predictions from analytical Gibbs calculations. For stronger system-reservoir couplings, temperature gradients lead to substantial nonequilibrium effects in the correlation patterns. Beyond the scope of the present study is the extension to time-dependent driving to follow protocols to operate quantum heat engines also for stronger dissipation and in proximity to a possible quantum speed limit.

\section{Open-system chain model and stochastic equivalent}
\subsection{Model}

\begin{figure}[]
\begin{center}
\includegraphics[width=8cm]{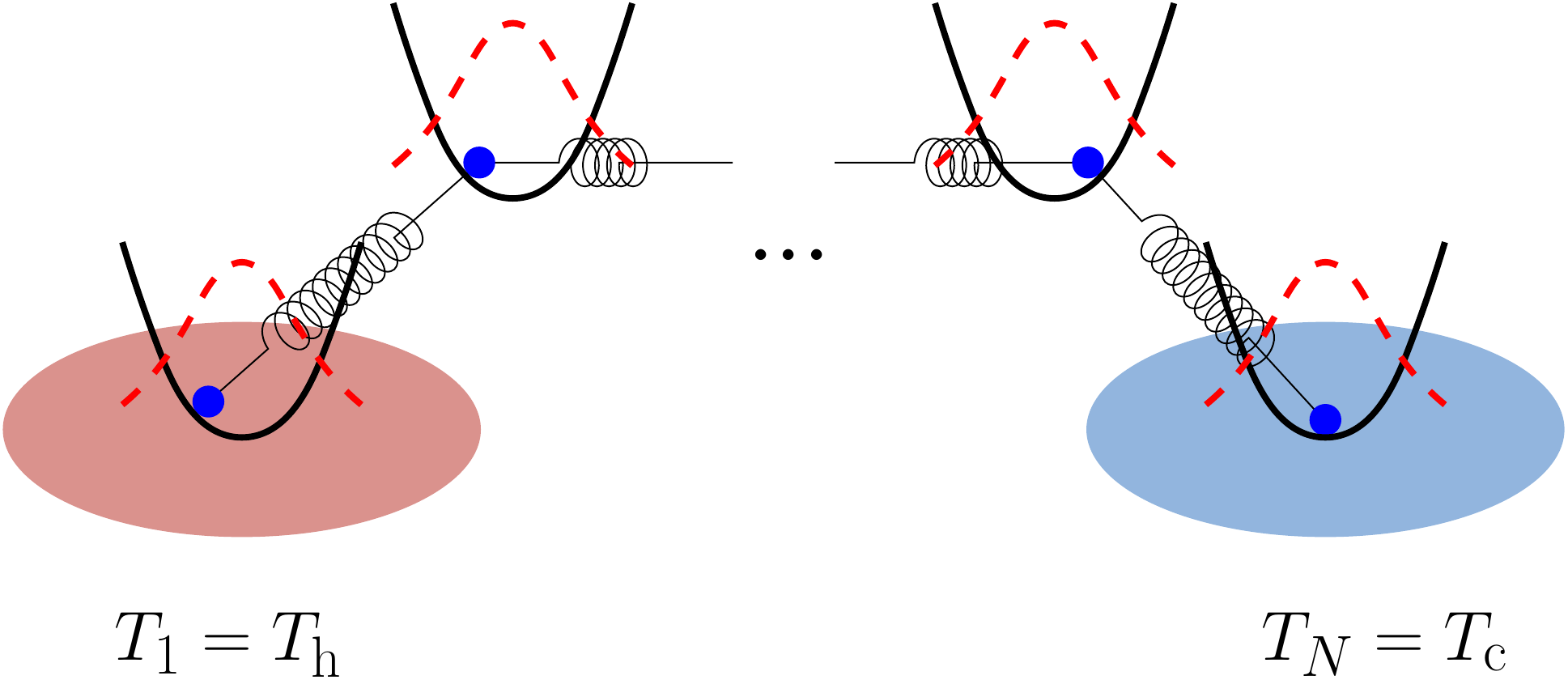}
\end{center}
\caption{Schematic of the investigated model: A chain of $N$ quadratically coupled oscillators which are connected to heat baths with temperatures $T_1=T_h$ and $T_N=T_c$ at the endpoints $n=1$ and $n=N$. The coupling strength between the oscillators is $\mu$, the on-site frequency $\omega_0$, the oscillator mass $m$ and the damping strengths caused by the coupling of the endpoints to the reservoirs are $\gamma_1=\gamma_N=\gamma$.}
\label{fig:osci_chain}
\end{figure}

A safe starting point to treat open quantum systems is to consider the full Hamiltonian of the global system where energy is conserved. This Hamiltonian has three parts which reflect system, coupling and reservoir, respectively,
\begin{equation}
H = H_s+\opind{H}{I}+\opind{H}{R}\;.
\label{eq:Ham_all}
\end{equation}
The system Hamiltonian is composed out of $N$ oscillators coupled by a quadratic nearest-neighbor two-body potential with strength $\mu$:

%
\begin{equation}
H_s = \sum_{n=1}^N\frac{p_n^2}{2m}+\frac{1}{2}m\omega_0^2 q_n^2+\frac{\mu}{2}\sum_{n=1}^{N-1}(q_n-q_{n+1})^2\;.
\label{eq:Ham_qu}
\end{equation}
The reservoir has the usual form that is used to model quantum Brownian motion of a particle coupled to a thermal heat bath; a collection of harmonic oscillators with positions $x_{\alpha}$, momenta  $p_{\alpha}$ and masses $m_{\alpha}$ forming a quasi-continuum of reservoir modes with frequencies $\omega_{\alpha}$ ($\alpha$, $\alpha'$ for  reservoirs coupled to $q_1$ and $q_N$, respectively) \cite{Caldeira1983a}:
\begin{equation}
\opind{H}{R} = \sum_{\alpha}\frac{p^2_\alpha}{2m_\alpha} + \frac{m_\alpha\omega_\alpha^2}{2}x^2_\alpha + \sum_{\alpha'}\frac{p^2_{\alpha'}}{2m_{\alpha'}} + \frac{m_{\alpha'}\omega_{\alpha'}^2}{2}x^2_{\alpha'}\,.
\label{eq:HR}
\end{equation}
We assume that the first and the last oscillators with index $n=1$ and $n=N$ are coupled to a thermal reservoir (see figure~\ref{fig:osci_chain}). The coupling to the baths with the strengths $c_{\alpha}$ is bilinear in the positions of the system oscillators and reservoir coordinates 
\begin{equation}
\opind{H}{I} = q_1\sum_{\alpha}c_\alpha x_\alpha + q_N\sum_{\alpha'}c_{\alpha'} x_{\alpha'} +q_1^2\sum_{\alpha}\frac{c_\alpha^2}{2m_\alpha\omega_\alpha^2} + q_N^2\sum_{\alpha'}\frac{c_{\alpha'}^2}{2m_{\alpha'}\omega_{\alpha'}^2}
\end{equation}
with two additional quadratic potential terms (``counterterms''). These ensure that the coupling to the reservoir induces no net potential when the reservoir is eliminated adiabatically. In the classical limit, this model corresponds to a purely velocity-dependent friction force.

One can show that the reduced system dynamics depends on the microscopic parameters in~(\ref{eq:HR}) only in an aggregate form, which is commonly denoted by a spectral density \cite{Weiss}
\begin{equation}
J_r(\omega) = \frac{\pi}{2} \sum_\alpha \frac{c_\alpha^2}{m_\alpha \omega_\alpha}
\delta(\omega-\omega_\alpha).
\end{equation}
Indexing each reservoir by the chain element $r\in\{1,N\}$ it is coupled to, all dissipative effects can be described by the reservoir correlation functions
\begin{equation}
L_r(t) = \frac{\hbar}{\pi}\int_0^\infty\rmd \omega J_r(\omega)[\coth\Big(\frac{\beta_r\hbar\omega}{2}\Big)\cos(\omega t) - \rmi\sin(\omega t)]
\label{eq:bath_corr}
\end{equation}
that depend on the respective spectral density and the inverse temperature $\beta_r=1/\kb T_r$.

For the spectral density of our model we choose $J_r(\omega)=m\gamma\omega(1+(\omega/\omega_c)^2)^{-2}$ with a UV cut-off frequency $\omega_c$. This constitutes an Ohmic reservoir with damping strength $\gamma_1=\gamma_N=\gamma$ for frequencies far below the cut-off $\omega_c$. With the choice of the spectral density and temperature, the thermally averaged motion of the chain is determined as a reduced dynamics by formally tracing out the reservoir. The imaginary part in~(\ref{eq:bath_corr}) 
arises from time ordering; Weyl ordering yields only the real part.

This is sufficient when expectation values of energy currents are computed, as, e.g., in \cite{Gaul2007}, but the imaginary part is essential when current correlations are considered, which are given by fourth-order correlations of positions and momenta. Since the decomposition of higher-order correlations using Wick's theorem relies on time-ordered two-time correlations, the full complex expression~(\ref{eq:bath_corr}) will be needed when we consider current-current correlations in section \ref{sec:curcurcorr}.

\subsection{Reduced dynamics and stochastic mapping}
The global Hamiltonian~(\ref{eq:Ham_all}) uniquely defines the dynamics of the global density operator in terms of the Liouville-von Neumann equation. Since this is a high-dimensional system, a reduced dynamics in terms of the system density operator is needed.
However, defining a reduced dynamics fully consistent with the global dynamics is difficult, particularly when the damping is large enough to modify the reduced equilibrium state or when the system is too complex for dissipation to be analyzed in terms of the system's level structure.

Path integrals involving a Feynman-Vernon influence functional \cite{FeynmanVernon1963} provide an exact approach in these settings. However, they cannot easily be transformed into an equivalent equation of motion.
The influence functional of a thermal oscillator bath is a Gaussian functional of the double path representing the propagation of the reduced density matrix. This property can be used to construct the functional as the stochastic average of an exponentiated action functional corresponding to external driving by colored Gaussian $c$-number noise.
A corresponding unraveling procedure \cite{Stockburger1998,stock02} of the functional employs two Gaussian processes whose correlation matrix reproduces both the real and imaginary parts of the reservoir correlation function~(\ref{eq:bath_corr}). This translates into a time-local stochastic equation of motion for the reduced density called the stochastic Liouville-von Neumann equation (SLN). This equation is formally exact and has proven to be useful for damped systems exposed to strong driving \cite{Schmidt2011, Schmidt2013} and energy transfer in systems coupled to reservoirs with structured spectral density \cite{Imai2015}.


If the spectral density is Ohmic, the imaginary part in $L_r(t)$ reduces to a derivative of a delta-function and therefore is time-local. Then, only the real part in ~(\ref{eq:bath_corr}) needs to be reconstructed by a non-Markovian noise term. The SLN then reduces to the so called stochastic Liouville-von Neumann equation for dissipation (SLED) \cite{Stockburger1999}:

\begin{eqnarray}
\difft\rho_\xi = \mathcal{L} \rho_\xi &=&-\frac{\rmi}{\hbar}[H_s,\rho_\xi] + \frac{\rmi}{\hbar}\xi_1(t)[q_1,\rho_\xi]+ \frac{\rmi}{\hbar}\xi_N(t)[q_N,\rho_\xi]\nonumber\\
&&-\frac{\rmi}{\hbar}\frac{\gamma_1}{2}[q_1,\{p_1,\rho_\xi\}]-\frac{\rmi}{\hbar}\frac{\gamma_N}{2}[q_N,\{p_N,\rho_\xi\}]\nonumber\;,
\label{eq:SLED}
\end{eqnarray}
with the damping constants $\gamma$, providing the coupling of the chain's endpoints to the baths, and the quantum noise $\xi_{r}(t)$ whose correlation function is given by the real part of $L_r(t-t')$
\begin{equation}
\langle\xi_r(t)\xi_{r'}(t')\rangle = \delta_{r,r'}\Re L_r(t-t')\;,
\label{eq:noisecor}
\end{equation}
where two distinct reservoirs, indexed with $r$ and $r'$, act independently on the system. We emphasize that even at $\kb T=0$ quantum noise is still present since the $\coth$-function in (\ref{eq:bath_corr}) does not vanish in this limit. Gardiner has identified a similar equation as an adjoint equation \cite{gardi88} of the quantum Langevin~\cite{Ford1965}. In the present context, we consider the Schr\"odinger picture the primary formalism of the dynamics; the adjoint dynamics introduced later in the present work will propagate time-dependent observables.
The physical reduced density matrix $\rho$ of the system is given by the expectation value of the stochastic density $\rho = \langle\opind{\rho}{\xi}\opind{\rangle}{stoch}$.

\subsection{Parameterization through cumulants}
For a harmonic system, SLED dynamics leads to Gaussian states for long enough times, both for individual samples $\rho_\xi$ and the physical density matrix even when the initial state is not Gaussian. We restrict ourselves to Gaussian states in the following and characterize them through first and second cumulants (expectation values and covariances) of positions and momenta.

Moments of observables are obtained from $\rho_\xi$ by a \emph{double} average, the combination of a trace operation $\langle \cdot\opind{\rangle}{tr} = \tr(\cdot\rho_\xi)$ and an expectation value with respect to the noise statistics,
\begin{equation}
\llangle A\rrangle = \llangle A\opind{\rangle}{tr}\opind{\rangle}{stoch}\;.
\label{eq:doubleav}
\end{equation}
Since we are dealing with a linear system, the twice-averaged first moments show effective classical behaviour, i.e., exponentially damped oscillations. We eliminate these by choosing their initial values to be zero.

For the covariance of two arbitrary operators $A$ and $B$, the double average allows the transformation
\begin{eqnarray}
\mathrm{Cov}(A,B) &  = & \textstyle{\frac{1}{2}}\llangle AB+BA\rrangle - \llangle A\rrangle\llangle B\rrangle\nonumber\\
& = & \textstyle{\frac{1}{2}}\llangle AB+ BA\rrangle - \llangle A\opind{\rangle}{tr}\langle B\opind{\rangle}{tr}\opind{\rangle}{stoch}
\nonumber\\
&&+ \llangle A\opind{\rangle}{tr}\langle B\opind{\rangle}{tr}\opind{\rangle}{stoch} - \llangle A\rrangle\llangle B\rrangle\\
& = & \langle\mathrm{Cov}_\mathrm{tr}(A,B)\opind{\rangle}{stoch} + \mathrm{Cov}_\mathrm{stoch}(\langle A\opind{\rangle}{tr},\langle B\opind{\rangle}{tr})
\label{eq:covsplit}
\end{eqnarray}
which effectively splits the covariance into two terms which we will call mean trace covariance and stochastic covariance.
It will be advantageous to treat these separately:
When mean trace covariance (mtr) and stochastic covariance (msc) are considered with $A$ and $B$ among all the coordinates of the operator-valued vector $\vec{\sigma}=(q_1,p_1,\dots,q_N,p_N)^t$, the covariance matrix $\BSig$ of its components $\sigma_j$ thus can be split as
\begin{eqnarray}
\BSig &=& \upperind{\BSig}{mtr} + \upperind{\BSig}{msc}\;,\\
\upperind{\BSig}{mtr}_{jk} &=&
\langle\mathrm{Cov}_\mathrm{tr}(\sigma_j,\sigma_k)\opind{\rangle}{stoch}\;,\\
\upperind{\BSig}{msc}_{jk} &=&
\mathrm{Cov}_\mathrm{stoch}(\langle \sigma_j\opind{\rangle}{tr},\langle \sigma_k\opind{\rangle}{tr})
\end{eqnarray}
This split will allow the translation of the SLED dynamics into \emph{deterministic} equations of motion for each of the two terms.

\subsection{Time evolution of system-trace cumulants}
As a first step toward equations of motion for $\upperind{\BSig}{mtr}$ and $\upperind{\BSig}{msc}$, we consider the time dependence of observables which are only averaged through $\langle \cdot \rangle_\mathrm{tr}$, i.e., quantities which are still random variables in the probability space of the Gaussian noise $\xi(t)$.

Their time evolution can be obtained from the adjoint dynamics of observables
associated with SLED.  Quite generally, this ``Heisenberg picture'' of a
quantum master equation is non-unitary equation of motion, governed by
the adjoint generator $\mathcal{L}^\dagger$ as described by Breuer and Petruccione \cite{Breuer}. It is not identical with the standard Heisenberg picture governed by the global Hamiltonian.

In the case of SLED, given by (\ref{eq:SLED}), the adjoint equation is a \emph{stochastic} equation of motion. The random adjoint-propagated operators $A_\xi(t)$ follow
\begin{eqnarray}
\difft A_\xi = & \frac{\rmi}{\hbar}[H_s,A_\xi] - \frac{\rmi}{\hbar}\xi_1(t)[q_1,A_\xi] - \frac{\rmi}{\hbar}\xi_N(t)[q_N,A_\xi]\nonumber\\
& + \frac{\rmi}{\hbar}\frac{\gamma_1}{2}\{p_1,[q_1,A_\xi]\} + \frac{\rmi}{\hbar}\frac{\gamma_N}{2}\{p_N,[q_N,A_\xi]\}\,.
\label{eq:adjoint_master}
\end{eqnarray}
When positions or momenta are inserted for $A_\xi$, this equation appears to be identical to the operator-valued quantum Langevin equation \cite{Ford1965,gardi88} at first glance. The subtle difference between the two approaches is the meaning of $\xi$. In the case of the quantum Langevin equation it is operator-valued; in our case it is real-valued. This eliminates certain ordering problems, a feature which will be helpful when evaluating higher-order correlations.
The price we pay for this lies in the fact that the operator algebra of canonical variables is lost in this propagation; the time evolution of products or functions of the canonical variables must be obtained separately by inserting them directly into~(\ref{eq:adjoint_master}).

It is helpful to note that the contributions in $\mathcal{L}^\dagger$ which provide the noises $\xi_r(t)$ are linear in $q_r$ while the others are all quadratic or bilinear in momentum and position. These contributions reduce to linear terms in the equations of motion for single coordinates, while the noise terms reduce to inhomogeneities $\xi_r(t)$. We therefore get a closed system of equations for the positions and momenta of the chain, which holds equally for first moments with respect to trace. These are shown in \ref{app:cumulants} and can be summarized 
in the linear equation
\begin{equation}
\difft\langle\vec{\sigma}\rangle_\mathrm{tr}=\mathbf{M}\langle\vec{\sigma}\rangle_\mathrm{tr}+\vec{\xi}(t)\,,
\label{eq:1st_cumulants}
\end{equation}
where $\vec{\xi}=(0, \xi_1, 0, \dots, 0, \xi_N)^t$ and where $\mathbf{M}$ is a $2N\times2N$ matrix which is determined by the parameters of the system Hamiltonian.

The evolution of the second cumulants is accessible with the adjoint SLED in a similar manner. Inserting $A=\sigma_j\sigma_k$ into~(\ref{eq:adjoint_master}) yields a right-hand side which is a linear combination of similar coordinate products, and products of the form $\sigma_l \xi_m$.
Combining the results for linear and quadratic terms yields the time derivatives of trace covariances, $\difft\,\mathrm{Cov}_\mathrm{tr}(\sigma_j,\sigma_k) = \difft
(\langle \sigma_j\sigma_k + \sigma_k\sigma_j\rangle_\mathrm{tr}/2 - \langle \sigma_j\rangle_\mathrm{tr}\langle \sigma_k\rangle_\mathrm{tr})$. The covariance matrix $\upperind{\BSig}{tr}
$ with elements $\mathrm{Cov}_\mathrm{tr}(\sigma_j,\sigma_k)$ obeys the simple equation
\begin{equation}
\difft \upperind{\BSig}{tr} = \mathbf{M}\upperind{\BSig}{tr}  + \upperind{\BSig}{tr} \mathbf{M}^{\dagger}\;.
\label{eq:2nd_cumulants}
\end{equation}
It is independent of $\vec{\xi}(t)$, reflecting the fact that a spatially homogeneous force cannot induce squeezing in this model.

\section{Deterministic evolution of the split covariance terms}
\label{sec:detev}

Due to the absence of noise in~(\ref{eq:2nd_cumulants}), $\upperind{\BSig}{tr}$ can immediately be identified with the mean trace covariance matrix $\upperind{\BSig}{mtr}$, i.e, the mean trace covariance can be computed by propagating
\begin{equation}
\difft \upperind{\BSig}{mtr} = \mathbf{M}\upperind{\BSig}{mtr}  + \upperind{\BSig}{mtr} \mathbf{M}^{\dagger}\;.
\label{eq:meantraceDGL}
\end{equation}

The stochastic covariance part simplifies, under the initial conditions we have assumed, as
\begin{equation}
\upperind{\BSig}{msc}_{j,k} = \llangle\sigma_j  \opind{\rangle}{tr}\langle\sigma_k\opind{\rangle}{tr}\opind{\rangle}{stoch}\;.
\end{equation}
Inserting the formal solution of~(\ref{eq:1st_cumulants}),
\begin{equation}
\langle\vec{\sigma}\rangle_\mathrm{tr} = \int_0^t\rmd t'\mathbf{G}(t-t')\vec{\xi}(t')\;,
\label{eq:formsol}
\end{equation}
using Green's function, we obtain
\begin{equation}
\upperind{\BSig}{msc} =\int_0^t\rmd t'\int_0^t\rmd t'' \mathbf{G}(t-t')\langle\vec{\xi}(t')\vec{\xi}^t(t'')\opind{\rangle}{stoch}\mathbf{G}^\dagger (t-t'')
\label{eq:signoiseint}
\end{equation}
which can be evaluated without sampling over noise. The correlation
matrix $\langle\vec{\xi}(t')\vec{\xi}^t(t'')\opind{\rangle}{stoch}$
can easily be computed and tabulated from~(\ref{eq:noisecor})
by numerical Fourier transform or summation over Matsubara
frequencies. The only non-zero entries in this matrix are the second
and last diagonal elements, which contain the autocorrelation
functions of the two reservoirs.
In the stationary limit, (\ref{eq:signoiseint}) arises from both the present method and the quantum Langevin approach \cite{Gaul2007,Freitas2014}. 
When also incorporating the contribution of $\upperind{\BSig}{mtr}$, the present approach is exact at any timescale.

Instead of computing this double convolution integral through a normal mode analysis, we construct a formal dynamical system of modest size which contains the elements of $\upperind{\BSig}{msc}$ as dynamical variables. Performing the derivative with respect to time and introducing an auxiliary variable $\mathbf{y}$ yields a closed system of linear differential equations

\begin{equation}
\upperind{\dot{\BSig}}{msc}(t) =  \mathbf{y}^\dagger(t) + \mathbf{y}(t) + \mathbf{M}\upperind{\BSig}{msc}(t) + \upperind{\BSig}{msc}(t)\mathbf{M}^\dagger\label{eq:sigma_noise}
\end{equation}
\begin{equation}
\dot{\mathbf{y}}(t) = \mathbf{G}(t)\mathbf{L}(t) \label{eq:yfuncm}
\end{equation}
\begin{equation}
\dot{\mathbf{G}}(t) =\mathbf{M}\mathbf{G}(t)\,,\label{eq:greenfunc}
\end{equation}
with the matrix $\mathbf{L}(t) = \langle\vec{\xi}(t)\vec{\xi}^t(0)\opind{\rangle}{stoch}$. As initial conditions 
we have $\mathbf{G}(0)=\mathbb{1}$, and we choose throughout the whole paper $\upperind{\BSig}{msc}(0) = 0$ as well as
${\mathbf{y}}(0)= 0$ (no initial system-reservoir correlations). $\upperind{\BSig}{mtr}(0)$ is determined by the unperturbed ground states of the chain's oscillators. The solutions from~(\ref{eq:2nd_cumulants}) and~(\ref{eq:sigma_noise})--(\ref{eq:greenfunc}) enable one to exactly calculate the transient dynamics of $\BSig$ which provides all information about the state of the system. Since $\mathbf{M}$ is taken from a dissipative linear system, the propagation of these equations is numerically stable even for large chain length.
\begin{figure*}[]
\begin{minipage}{8cm}
\includegraphics[width=7.6cm]{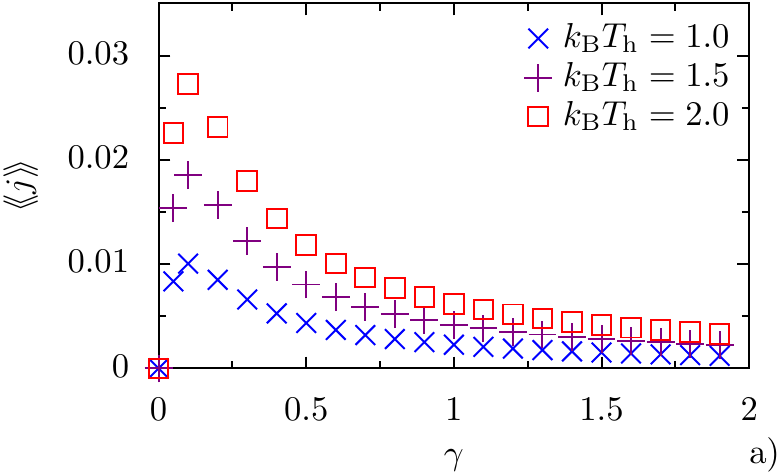}
\end{minipage}
\hfill
\begin{minipage}{8cm}
\includegraphics[width=7.6cm]{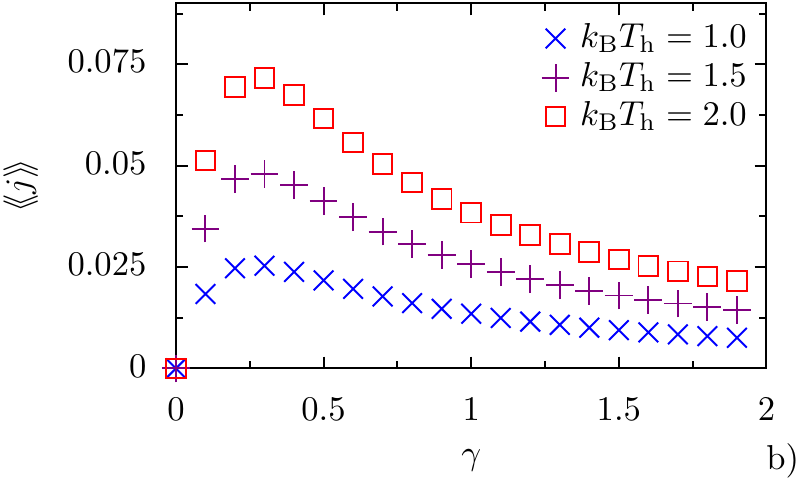}
\end{minipage}
\caption{Steady-state heat flux $\llangle j\rrangle$ versus the damping strength $\gamma$ for a system as shown in figure~\ref{fig:osci_chain} with $N=20$ oscillators and $\gamma$ denoting the damping of the chain's endpoints at $n=1$ and $n=N$. The different colors show results for varying temperatures of the hot reservoir while the cold reservoir remains at $T_\mathrm{c}=0$; thus this bath causes only quantum fluctuations. The couplings between the oscillators are $\mu=0.1$ (a) and $\mu=0.3$ (b) leading to a positive shift in the current's maximum for larger $\mu$. Other parameters are  $m=1.0$ and $\omega_0=1.0$.}
\label{fig:j_over_eta}
\end{figure*}
For an efficient computation of steady-state behavior, we set the derivative with respect to time in~(\ref{eq:sigma_noise}) equal to zero and obtain a Lyapunov equation
\begin{equation}
\mathbf{M}\upperind{\BSig}{msc} + \upperind{\BSig}{msc}\mathbf{M}^\dagger\  + \mathbf{y}^\dagger + \mathbf{y} = 0
\label{eq:sigma_noisesteady}
\end{equation}
with an inhomogeneity that is determined by integrating (\ref{eq:yfuncm}) and (\ref{eq:greenfunc}) to times, large enough that $\mathbf{y}$ reaches a constant steady-state value.

A physical interpretation of $\mathbf{y}$ is manifested by integration of~(\ref{eq:yfuncm})
\begin{equation}
\mathbf{y}(t) = \mathbf{y}(0) + \int_0^t\rmd t' \mathbf{G}(t-t')\langle\vec{\xi}(t)\vec{\xi}^t(t')\opind{\rangle}{stoch}\;.
\label{eq:yfunc_int}
\end{equation}
By comparison with~(\ref{eq:formsol}) this can be identified  as the correlations of the bath fluctuations with the system's degrees of freedom given by first cumulants:
\begin{equation}
\mathbf{y}(t) = \llangle\vec{\sigma}\opind{\rangle}{tr}(t)\vec{\xi}^t(t)\opind{\rangle}{stoch}\,,
\label{eq:sysbathcorr}
\end{equation}
where $\mathbf{y}(0)=0$. With this interpretation, $\xi_r(t)$ appears as a `stochastic substitute' for a reservoir operator. The time until $\mathbf{y}(t)$ is in equilibrium is provided by the time until the integrand in~(\ref{eq:yfunc_int}) decays to zero. So for moderate damping, the correlation time in $\mathbf{L}(t-t')$ is the pivotal time scale.

\section{Energy flux operators}

With the approach we presented in the previous section we are able to determine the state of a harmonic quantum chain coupled to Ohmic reservoirs with arbitrary temperatures. This allows us to study nonequilibrium situations with finite heat fluxes causing an energy transport from a hot to a cold reservoir. The link topology of the chain shown in figure~\ref{fig:osci_chain} suggests to consider three cases. Two of them account for the coupling of the endpoints to the neighbor and hot or cold reservoir, respectively, while the third covers the oscillators in the bulk which are coupled to next neighbors. By employing $\mathcal{L}^\dagger$ on $H_n = \frac{p_n^2}{2m}+\frac{1}{2}m\omega_0^2 q_n^2+\frac{\mu}{2}[(q_{n-1}-q_n)^2 + (q_n-q_{n+1})^2]$, we obtain three adjoint dynamical equations for the energy operators:

\begin{eqnarray}
&\difft H_1 = \frac{1}{m}\xi_1(t) p_1 - \gamma_1\frac{p_1^2}{m} + \frac{\mu}{m} q_{2}p_1\nonumber\\
&\difft H_n =\frac{\mu}{m} q_{n-1}p_n + \frac{\mu}{m} q_{n+1}p_n,\quad 2\leq n\leq N-1\nonumber\\
&\difft H_N = \frac{\mu}{m} q_{N-1}p_N + \frac{1}{m}\xi_N(t) p_N- \gamma_N\frac{p_N^2}{m}\,. \nonumber\\
\end{eqnarray}
For simplicity, the index $\xi$ has been omitted from the adjoint-propagated observables. The assumption of locality of energy transfer between nearest neighbors allows the identification of individual terms on the r.h.s. of $\difft H_n=j_{n-1,n}-j_{n,n+1}$, leading to
\begin{eqnarray}
j_{n-1,n} &= \frac{\mu}{m} q_{n-1}p_n,\quad\phantom{-} 2\leq n\leq N\nonumber\\
j_{n,n+1} &=  -\frac{\mu}{m} q_{n+1}p_n,\quad 1\leq n\leq N-1\,.
\end{eqnarray}
To get from the energy flux operators to the first moments of the heat flux, we have to apply the stochastic average on the trace averaged operators. Since we use initial conditions where all expectation values of phase-space operators are zero, the resulting currents are given by the elements of the covariance matrix:
\begin{eqnarray}
\llangle j_{n-1,n}\rrangle = \frac{\mu}{m}\llangle q_{n-1}p_n\rrangle,\quad\phantom{-} 2\leq n\leq N\label{eq:heatfluxleft}\\
\llangle j_{n,n+1}\rrangle =  -\frac{\mu}{m}\llangle q_{n+1}p_n\rrangle,\quad 1\leq n\leq N-1\label{eq:heatfluxright}\\
\llangle j_{r,1}\rrangle = \frac{1}{m}\langle\xi_1(t)\langle p_1\opind{\rangle}{tr}\opind{\rangle}{stoch} - \gamma_1\Big\langle\hspace{-3.5pt}\Big\langle\frac{p_1^2}{m}\Big\rangle\hspace{-3.5pt}\Big\rangle\label{eq:heatfluxres1}\\
\llangle j_{r,N}\rrangle = -\frac{1}{m}\langle\xi_N(t)\langle p_N\opind{\rangle}{tr}\opind{\rangle}{stoch} + \gamma_N\Big\langle\hspace{-3.5pt}\Big\langle\frac{p_N^2}{m}\Big\rangle\hspace{-3.5pt}\Big\rangle\label{eq:heatfluxresN}\,,
\end{eqnarray}
where  the second term in~(\ref{eq:heatfluxres1}) and (\ref{eq:heatfluxresN}) is classical while the first term reflects correlations of the bath fluctuations with the system's degrees of freedom. According to~(\ref{eq:sysbathcorr}), this term is an element of $\mathbf{y}$ which needs already to be computed for $\BSig$, and does not demand any further effort. In a steady state, the current in the bulk is constant over the chain and equal to the current between either bath and the system. Therefore, for the analysis of steady-state currents we write  $\llangle j_{r,1}\rrangle=\llangle j_{n,n+1}\rrangle=\llangle j\rrangle$.

The heat current provides us intuitive consistency checks such as a vanishing heat flux in the absence of a temperature gradient. This seemingly trivial result can not be reproduced by naively constructed local Lindblad operators as it was shown for harmonic oscillators and two level systems \cite{Levy2014, Stockburger2016}. To this respect, our approach delivers consistent results for various combinations of couplings and temperatures. Particularly also low temperatures and unequal damping strengths of the endpoints do not lead to a violation of the 2nd law.

An interesting feature of the steady-state flux that is shown in figure~\ref{fig:j_over_eta} is the interplay of the damping strengths $\gamma$ and the couplings $\mu$ within the oscillators. For small damping, the plots show a linear increase of $\llangle j\rrangle$ with the damping $\gamma$ up to a maximum followed by an algebraic decay according to $\propto\gamma^{-1}$ for large damping strengths. This behavior was studied by Rieder et al. in~\cite{Rieder1967} for a system of classical harmonic oscillators without local frequencies $\omega_0$ and by Gaul and B\"uttner \cite{Gaul2007} for a similar quantum chain as the one used here. We could verify the observation of~\cite{Gaul2007} that the position of the current's maximum increases linearly with the next neighbor coupling $\mu$ for small couplings and falls back for large $\mu$. The distinct sensitivity of the current on the damping will be relevant when we study heat-flux correlations, under equilibrium and nonequilibrium conditions in the following sections.																																																																																																								 

\section{Higher order correlations}
\label{sec:curcurcorr}
\subsection{Stochastic calculations}

The relative simplicity of the formal dynamics for the covariance matrix invites to study space-time correlations of heat flux expressed through the covariance. As the fluxes themselves are given by a correlation of phase-space variables (see~(\ref{eq:heatfluxleft}) and (\ref{eq:heatfluxright})), the heat-flux correlations represent an average over four operators which are simplified by an application of Wick's theorem reducing the Gaussian heat-flux correlations to products of phase-space correlations. 
The covariance of two currents $j_{n,n+1}(t)=j_n(t)$ and $j_{l,l+1}(t')=j_l(t')$ reads
\begin{equation}
\llangle j_{n}(t)j_{l}(t')\rrangle_c = \llangle j_{n}(t)j_{l}(t')\rrangle-\llangle j_{n}(t)\rrangle\llangle j_{l}(t')\rrangle\;,
\label{eq:covheat}
\end{equation}
with the last term being finite for nonequilibrium settings. Inserting~(\ref{eq:heatfluxright}) gives
\begin{eqnarray}
\llangle j_n(t)j_l(t')\rrangle_c =
\Big(\frac{\mu}{m}\Big)^2[&\llangle q_{n+1}(t)p_n(t)q_{l+1}(t')p_l(t')\rrangle\nonumber\\
&-\llangle q_{n+1}(t)p_n(t)\rrangle\llangle q_{l+1}(t')p_l(t')\rrangle]
\end{eqnarray}
where Wick's theorem decomposes the first term to products of phase-space variables

\begin{eqnarray}
\llangle j_n(t)j_l(t')\rrangle_c =
\Big(\frac{\mu}{m}\Big)^2[&\llangle q_{n+1}(t)q_{l+1}(t')\rrangle\llangle
p_n(t)p_l(t')\rrangle \nonumber\\
&+\llangle q_{n+1}(t)p_l(t')\rrangle\llangle p_n(t)q_{l+1}(t')\rrangle]\,.
\label{eq:covhettwo2}
\end{eqnarray}

Here we consider time-ordered products; operators appearing at equal time in~(\ref{eq:covhettwo2}) commute since they bear different site indices. For definiteness, we assume $t>t'$. Since we wish to consider these correlations in the stationary state, the initial preparation will not be at time $t_0 = 0$, but at time $t_0 = -\infty$ in the following. The time-ordered correlation matrix $\BSig_>(t-t')$ is then only a function of the time difference $\Delta = t-t'$. Its value at $\Delta = 0$ is related to the results of section \ref{sec:detev},
\begin{equation}
\BSig_>(0) = \lim_{\tau\to\infty} \BSig(\tau) + \BSig''_0
= \lim_{\tau\to\infty} \upperind{\BSig}{mtr} (\tau) + \BSig''_0
\label{eq:sigordini}
\end{equation}
where
\begin{equation}
\BSig''_0=
\left(\begin{array}{ccccc}
0 & \frac{\rmi}{2} & & &\\
-\frac{\rmi}{2} & 0 &\ddots & &\\
  &\ddots & \ddots &\ddots &\\
  & &\ddots & 0 & \frac{\rmi}{2}\\
  & & &-\frac{\rmi}{2}& 0
\end{array}\right)\,.
\label{eq:imsigmatwotinitial}
\end{equation}
reflects the usual commutation relations, with a sign convention which determines a convention for the ordering of positions and momenta at equal time.

For arbitrary $\Delta>0$, the elements of $\BSig_>(\Delta)$ are
correlation functions of the form
$\llangle A \mathcal{V}(\Delta,0) B \mathcal{V}(0,t_0)\rrangle$ with
\begin{equation}
\mathcal{V}(t_2,t_1) = \mathcal{T}_\leftarrow\exp\Big(\int_{t_1}^{t_2}\mathcal{L}(s)\rmd s\Big)\;.
\end{equation}
The derivative of these correlation functions with respect to $\Delta$ can be analyzed in terms of the adjoint propagation and cast into matrix-valued equations of motion as in section \ref{sec:detev}; the time derivative transforms as
\begin{eqnarray}
\frac{\partial}{\partial \Delta} \llangle A \mathcal{V}(\Delta,0) B \mathcal{V}(0,t_0)\rrangle
&=&
\llangle A \mathcal{L}\mathcal{V}(\Delta,0) B \mathcal{V}(0,t_0)\rrangle
\nonumber\\
&=&
\llangle \mathcal{L}^\dagger A \mathcal{V}(\Delta,0) B \mathcal{V}(0,t_0)\rrangle\,.
\label{eq:adjcorr}
\end{eqnarray}
In \ref{app:Corrs} we show that this translates into a system of equations of motion which can be summarized in matrix form as
\begin{eqnarray}
\frac{\partial}{\partial\Delta}\BSig_>(\Delta) &=& \mathbf{M}\BSig_>(\Delta) + \mathbf{z}(\Delta)\label{eq:corrpropsigma}\\
\frac{\partial}{\partial\Delta}\mathbf{z}(\Delta) &=& - \mathbf{L}(\Delta) - \mathbf{z}(\Delta) \mathbf{M}^\dagger
\label{eq:corrpropz}
\end{eqnarray}
with initial conditions given by~(\ref{eq:sigordini}) and by $\mathbf{z}(0) = \lim_{\tau\to\infty} \mathbf{y}^\dagger(\tau)$.

Like the results for the states shown in the previous sections, the correlations presented here are valid for arbitrary damping and temperatures. Therefore, our results overcome the restriction of weak damping  that has to be respected when applying the quantum regression hypothesis which is commonly used in models for quantum optics \cite{Carmichael2002}. It was pointed out that the hypothesis which is based on a Born-Markov approximation is violating fundamental consistency criteria such as the fluctuation-dissipation theorem \cite{Ford1996, Talkner1986, Callen1951}. The formal reason for the validity of~(\ref{eq:corrpropsigma}) and (\ref{eq:corrpropz}) lies in the fact that the propagation of the inhomogeneity $\mathbf{z}(\Delta)$ as well as the propagations leading to its initial value take into account fluctuations up to arbitrarily high order, which reflects the system-reservoir correlations also for strong damping and ensures the accordance with the fluctuation-dissipation theorem.

With the real- and imaginary parts of the phase-space correlations given, one can construct the real- and imaginary parts of the heat-flux correlations according to~(\ref{eq:covhettwo2}):

\begin{eqnarray}
\fl\mathrm{Re}\llangle j_n(t)j_l(t')\rrangle_c =
\Big(\frac{\mu}{m}\Big)^2&[\mathrm{Re}\llangle q_{n+1}q_{l+1}\rrangle\mathrm{Re}\llangle p_np_l\rrangle + \mathrm{Re}\llangle q_{n+1}p_l\rrangle\mathrm{Re}\llangle p_nq_{l+1}\rrangle\nonumber\\
&- \mathrm{Im}\llangle q_{n+1}q_{l+1}\rrangle\mathrm{Im}\llangle p_{n}p_l\rrangle - \mathrm{Im}\llangle q_{n+1}p_l\rrangle\mathrm{Im}\llangle p_nq_{l+1}\rrangle]\,,
\label{eq:real_covsled}
\end{eqnarray}
where we skipped the dependence of the correlations on $(t,\Delta)$ for a better legibility.
For the imaginary part one needs to compute

\begin{eqnarray}
\fl\mathrm{Im}\llangle j_n(t)j_l(t')\rrangle_c =
\Big(\frac{\mu}{m}\Big)^2&[\mathrm{Re}\llangle q_{n+1}q_{l+1}\rrangle \mathrm{Im}\llangle p_np_l\rrangle  + \mathrm{Re}\llangle q_{n+1}p_l\rrangle \mathrm{Im}\llangle p_nq_{l+1}\rrangle  \nonumber\\
&+ \mathrm{Im}\llangle q_{n+1}q_{l+1}\rrangle \mathrm{Re}\llangle p_np_l\rrangle  + \mathrm{Im}\llangle q_{n+1}p_l\rrangle \mathrm{Re}\llangle p_nq_{l+1}\rrangle ]\,.
\label{eq:im_covsled}
\end{eqnarray}
Again, one should note that the pairs of imaginary parts in~(\ref{eq:real_covsled}) provide contributions to the real parts of the heat-flux correlations.

\subsection{Current fluctuations in a Gibbs state}

\begin{figure*}[t]
\begin{minipage}{8cm}
\includegraphics[width=7.8cm]
{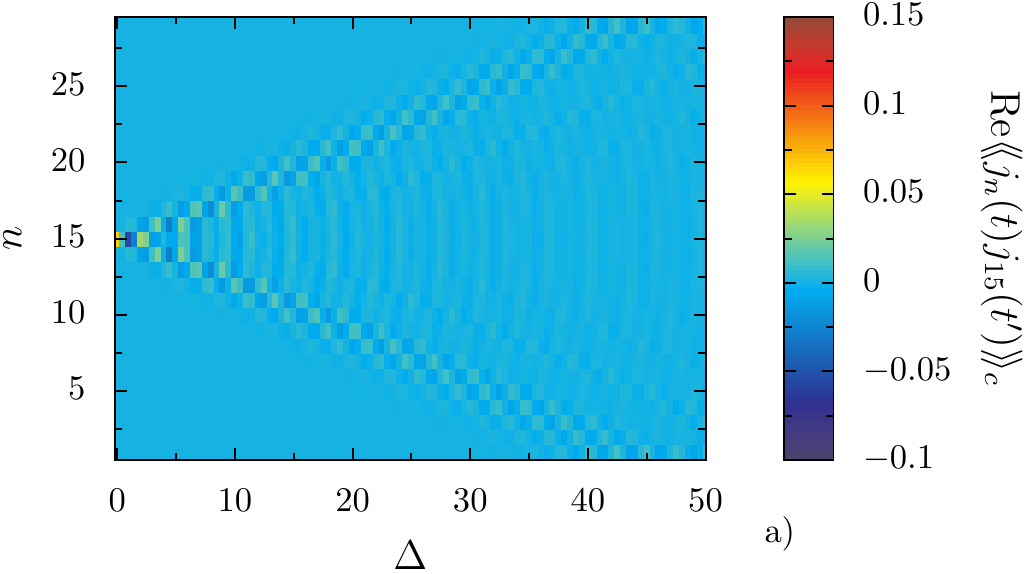}
\end{minipage}
\hfill
\begin{minipage}{8cm}
\includegraphics[width=7.8cm]{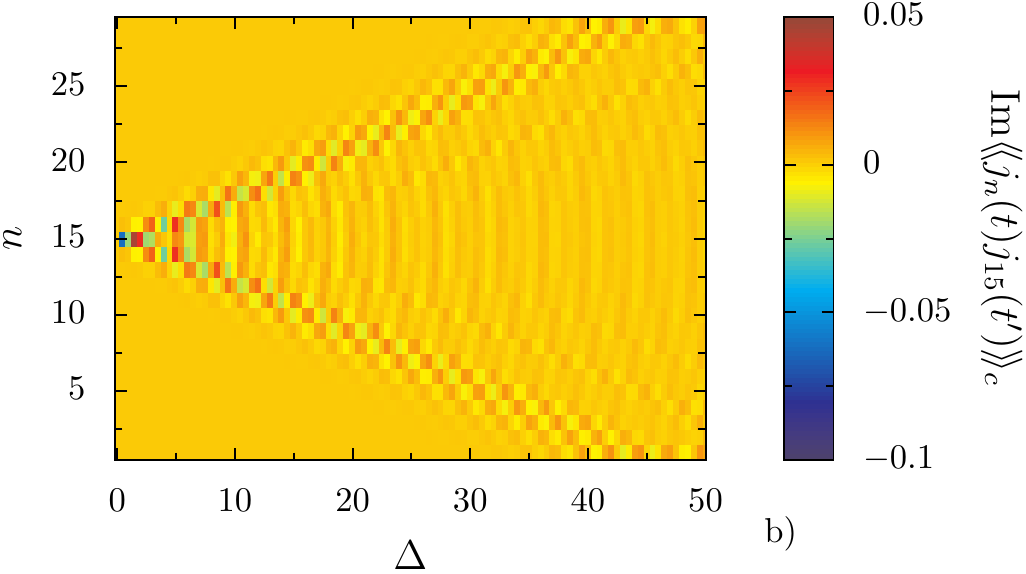}
\end{minipage}

\begin{minipage}{8cm}
\vspace{5pt}
\includegraphics[width=7.0cm]{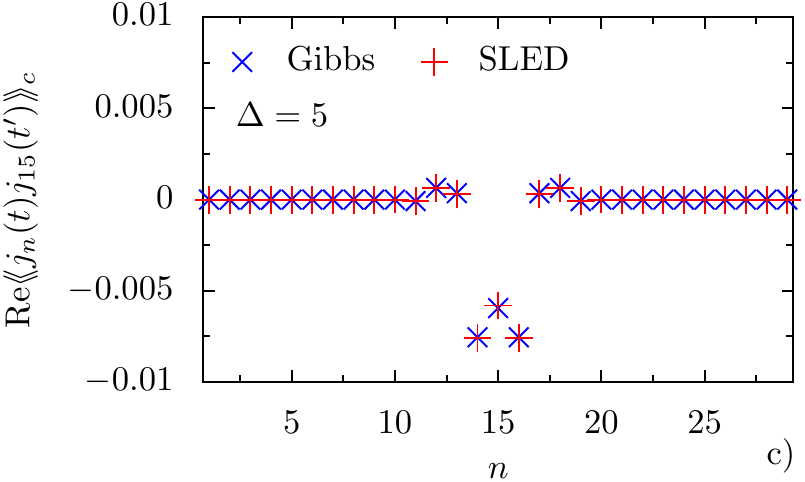}
\end{minipage}
\hfill
\begin{minipage}{8cm}
\vspace{5pt}
\includegraphics[width=7.0cm]{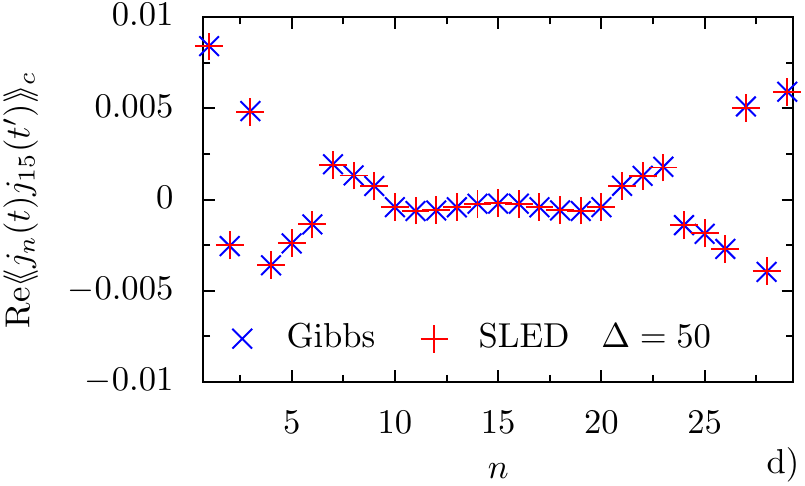}
\end{minipage}
\caption{The covariance $\llangle j_n(t)j_{15}(t')\rrangle_c$ from the SLED approach in a chain with the boundary conditions from~(\ref{eq:Ham_fixed}) and $N=30$ oscillators. The density plots a) and b) show the real and imaginary part of the covariance for varying time $\Delta=t-t'$ and index $n$. Plots c) and d) show $\mathrm{Re}\llangle j_n(t)j_{15}(t')\rrangle_c$ versus the index  $n$ for values of $\Delta=5$ (c) and $\Delta=50$ (d). Results for a canonical Gibbs ensemble are also shown in the lower plots. For the low damping of  $\gamma=10^{-4}$ used here, the results agree very well for all sites and long propagation times (d). Other parameters are $\kb T_r=0$, $\omega_0=1.0$, $m=1.0$ and $\mu=0.5$.}
\label{fig:SLED_Gibbs_comp_weak}
\end{figure*}

Finally, we perform analytic calculations which resemble the limit of zero damping ($\gamma=0$) and allow us to verify the validity of our results obtained by the previous calculations of heat-flux correlations. Therefore,
we use a Hamiltonian with fixed ends

\begin{equation}
H_s = \sum_{n=1}^N\frac{p_n^2}{2m}+\frac{1}{2}m\omega_0^2 q_n^2+\frac{\mu}{2}\sum_{n=1}^{N-1}(q_n-q_{n+1})^2 + \frac{\mu}{2}(q_1^2+q_N^2)\,,
\label{eq:Ham_fixed}
\end{equation}
which is an extension of~(\ref{eq:Ham_qu}) to boundary conditions that easily can be incorporated in the numerical SLED dynamics for comparison. The Hamiltonian in~(\ref{eq:Ham_fixed}) can be interpreted as a chain with $N+2$ oscillators which yield $q_0=q_{N+1}=0$. To achieve a normal mode representation of $H_s$, we introduce the transformations
\begin{equation}
q_n = \sqrt{\frac{2}{N+1}}\sum_{\nu=1}^N\sin(k_\nu n)\mathcal{Q}_\nu\qquad
p_n = \sqrt{\frac{2}{N+1}}\sum_{\nu=1}^N\sin(k_\nu n)\mathcal{P}_\nu
\label{eq:Trafos}
\end{equation}
with the wave vector of the $\nu$th normalmode $k_\nu = \pi \nu/(N+1),\;\nu = \mathbb{N}_0$.
Applying these transformations and exploiting the fixed boundary conditions we get:
\begin{equation}
H_s = \sum_\nu\Big[\frac{\mathcal{P}_\nu^2}{2m} + \frac{1}{2}m\omega^2(k_\nu)\mathcal{Q}^2_\nu\Big]\,.
\label{eq:Ham_kspace}
\end{equation}
This Hamiltonian constitutes a sum of uncoupled oscillators, each with frequency $\omega(k_\nu)=\sqrt{\omega_0^2 +(4\mu/m)\sin^2\Big(k_\nu/2\Big)}$. Applying the Hamilton equations gives us the standard differential equation of the harmonic oscillator for the $\nu$th mode and the classical dynamics of $\mathcal{Q}_\nu(\Delta)$ and $\mathcal{P}_\nu(\Delta)$ whose quantum nature is provided by the non-commutativity of the initial values $\mathcal{Q}_\nu(0)$ and $\mathcal{P}_\nu(0)$.
Therefore, thermal averages over pairs of the initial values have to be calculated and put together with the classical dynamics plus the transformations from~(\ref{eq:Trafos}) to arrive at

\begin{eqnarray}
\langle q_n(\Delta)q_l(0)\opind{\rangle}{tr}&=\frac{2}{N+1}\frac{1}{2m}\sum_\nu\sin(k_\nu n)\sin(k_\nu l)\nonumber\\
&\times\frac{1}{\omega(k_\nu)}[\coth\Big(\frac{\beta\omega(k_\nu)}{2}\Big)\cos(\omega(k_\nu)\Delta)-\rmi\sin(\omega(k_\nu)\Delta)]\,,
\label{eq:qq_gibbs}
\end{eqnarray}
where we only have to take the trace average of a canonical Gibbs state into account and can skip the stochastic average that occurs in the numerical method. From the correlation of positions, by considering $p_n(\Delta) = m\dot{q}_n(\Delta)$ one can calculate all other elements of the correlation matrix $\BSig(\Delta)$ from~(\ref{eq:covhettwo2}) and  the sought after higher order correlations.

Without the particular focus on correlations of heat currents, the present canonical calculation has been presented in \cite{Ford1965} under allusion to the extendability to any higher order correlation function with an even number of operators. In that paper and for example in \cite{Zurcher1990} the thermodynamic Gibbs calculation is compared to dynamical models treated with Langevin equations. But particularly multi-time correlations are challenging in this framework, as the operator valued quantum noise demands careful operator ordering.

\begin{figure*}[t]
\begin{minipage}{8cm}
\includegraphics[width=7.8cm]{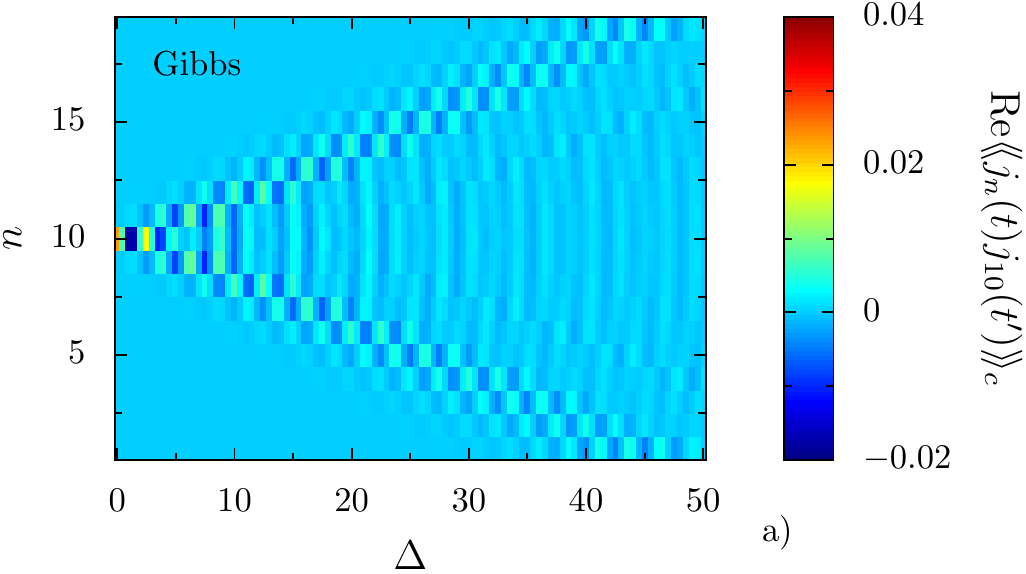}
\end{minipage}
\hfill
\begin{minipage}{8cm}
\includegraphics[width=7.8cm]{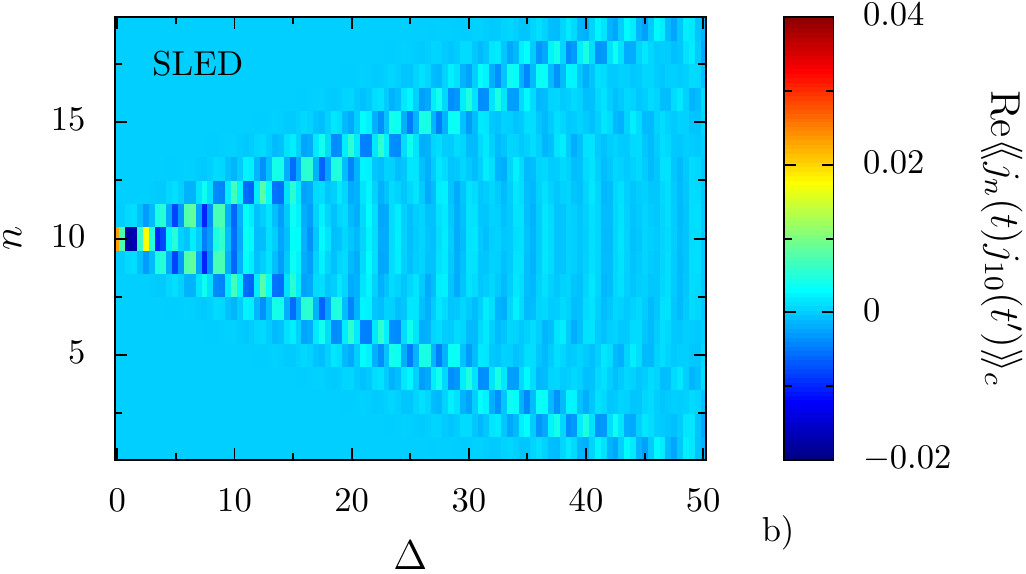}
\end{minipage}

\begin{minipage}{8cm}
\vspace{5pt}
\includegraphics[width=7.0cm]{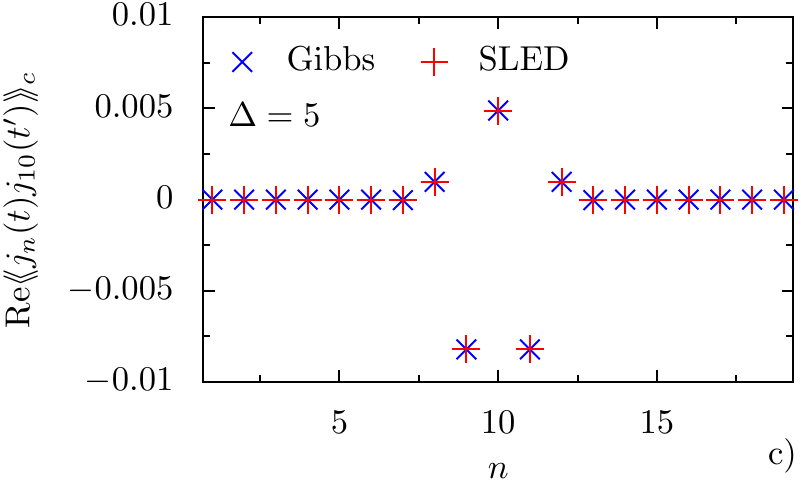}
\end{minipage}
\hfill
\begin{minipage}{8cm}
\vspace{5pt}
\includegraphics[width=7.0cm]{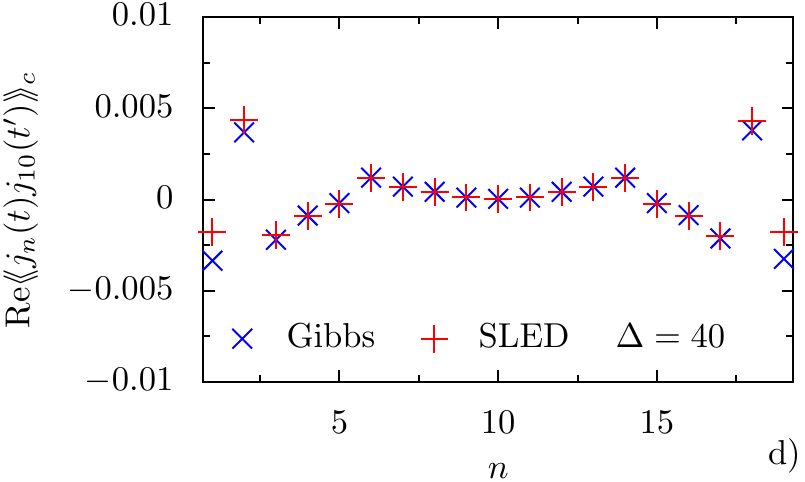}
\end{minipage}
\caption{Real part $\mathrm{Re}\llangle j_n(t)j_{10}(t')\rrangle_c$ of the covariance in a chain such as shown in figure~\ref{fig:osci_chain} with $N=20$ and equal temperature $\kb T_r=0$ in both reservoirs. Plot a) shows the space-time pattern calculated from Gibbs and b) with SLED for a damping of $\gamma=0.5$. The damping on the chain's endpoints leads to deviations from the Gibbs results for $\Delta$ larger than $\sim 40$. This is clear from plots c) and d) which show the covariance plotted over the index $n$ at $\Delta=5$ and $\Delta=40$. Other parameters are $\omega_0=1.0$, $m=1.0$ and $\mu=0.3$.}
\label{fig:SLED_Gibbs_comp_strong}
\end{figure*}

In order to check the validity of the dynamical description based on the SLED and our numerical implementation we compare its results with the ones from a thermal Gibbs state. We consider the Hamiltonian in~(\ref{eq:Ham_fixed}) and compute the covariance of heat fluxes for varying time difference $\Delta=t-t'$ and lead index $n$. Figure~\ref{fig:SLED_Gibbs_comp_weak} plots the real- and imaginary parts of $\llangle j_n(t)j_{N/2}(t')\rrangle_c$ in density plots (top) for a chain with $N=30$ oscillators where the ends are weakly damped with $\gamma=10^{-4}$ and the reservoir temperatures are equal with $\kb T_r=0$. For such a weak damping we expect a good agreement with the numerical method based on the SLED and the results of the Gibbs calculations, which represent the limit of zero damping. Indeed, a quantitative agreement is shown in the lower plots of figure~\ref{fig:SLED_Gibbs_comp_weak} where the covariance versus the index $n$ is shown for two different values of $\Delta$. For a small time difference $\Delta=5$ (c) the results of SLED and Gibbs agree with high accuracy, as well as for a larger $\Delta=50$ (d). Particularly the agreement of the two methods at the endpoints of the chain indicate that we are in a limit where damping is negligible over the considered time scales since one would expect that effects of damping  first start to emerge at the oscillators close to the reservoirs.

\section{Heat-flux correlations away from equilibrium}

If larger damping like $\gamma=0.5$ is considered in the SLED approach as done in figure~\ref{fig:SLED_Gibbs_comp_strong} one observes deviations of the covariances resulting from the stochastic approach and the assumption of a canonical Gibbs state. The upper plots show the real part of the spatiotemporal correlations from analytical Gibbs calculations (a) and numerical simulations (b). While the correlations are undamped for the canonical ensemble, the dynamics from the SLED are weakly damped  as apparent for time differences $\Delta$ which are large enough to find finite correlations at the chain's endpoints. This effect of damping is more pronounced in plots c) and d) showing the covariance over the index $n$ for two different times $\Delta$. While the left plot which shows  $\mathrm{Re}\llangle j_n(t)j_{N/2}(t')\rrangle_c$ for $\Delta=5$ reveals a good agreement for the whole chain, deviations of Gibbs and SLED become apparent in the vicinity of the endpoints if the time difference is larger like $\Delta= 40$ in the right plot. With the previous validity checks, we show that our numerical approach reflects the limit of zero damping of the canonical Gibbs ensemble while deviations for finite damping occur, as to be expected, at the ends of the chain which are coupled to reservoirs. We have also found agreement between the present methods for small dampings and finite but equal temperatures at the attached reservoirs. Nevertheless, nonequilibrium situations caused by different temperatures remain withheld to the numerical SLED method.

For the model with reservoirs being only attached at the endpoints of a sufficiently long chain, one has to admit that the damping of the reservoir only affects correlations in the vicinity of the end points. This might question the need of a numerical method valid for strong damping, as the Gibbs calculation presented here delivers correct results for the bulk. But a pitfall that arises particularly for large chains is the decreasing level spacing of the Hamiltonian with increasing number of modes. This induces rigid upper bounds for the damping to ensure that the reservoir induced level broadening is much smaller than the level spacing, which is important for all formalisms where the system is supposed to be in a Gibbs state. Besides the damping strength, we have also the freedom to vary the temperatures leading to a nonequilibrium situation. This can not be calculated by a canonical Gibbs state and provides us effects on the entire chain.

\begin{figure*}[t]
\begin{minipage}{8cm}
\includegraphics[width=7.8cm]{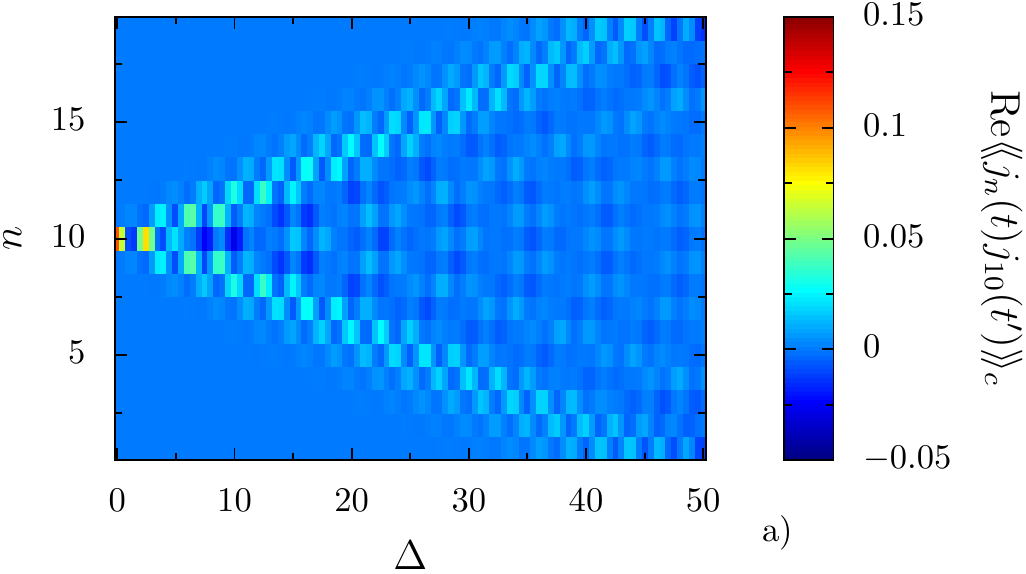}
\end{minipage}
\hfill
\begin{minipage}{8cm}
\includegraphics[width=7.8cm]{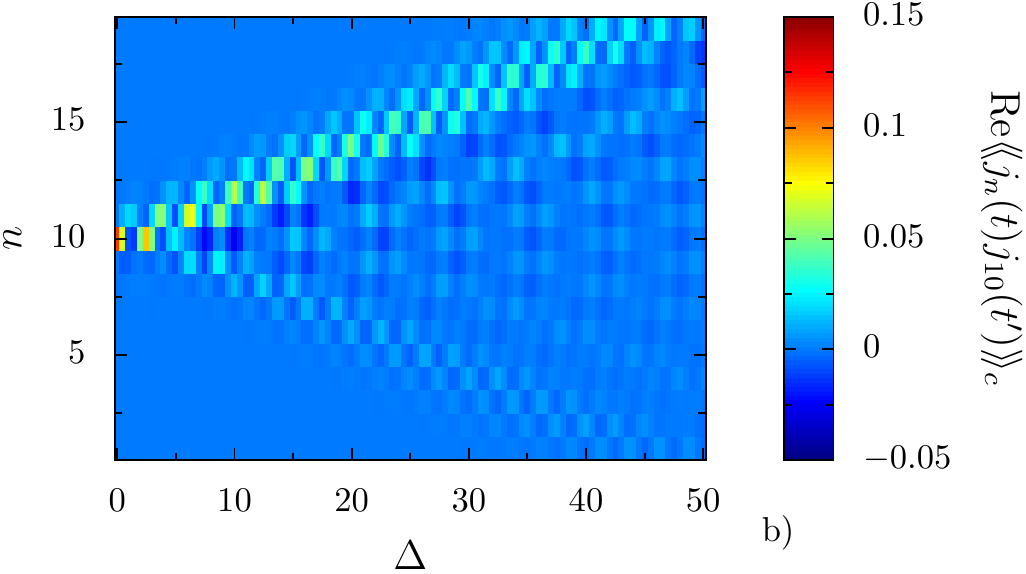}
\end{minipage}
\vspace{10pt}
\begin{minipage}{8cm}
\includegraphics[width=7.3cm]{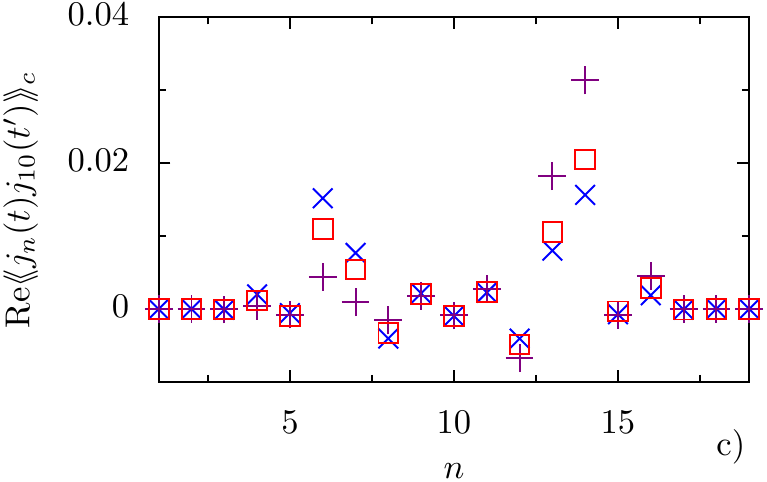}
\end{minipage}
\hfill
\begin{minipage}{8cm}
\includegraphics[width=7.8cm]{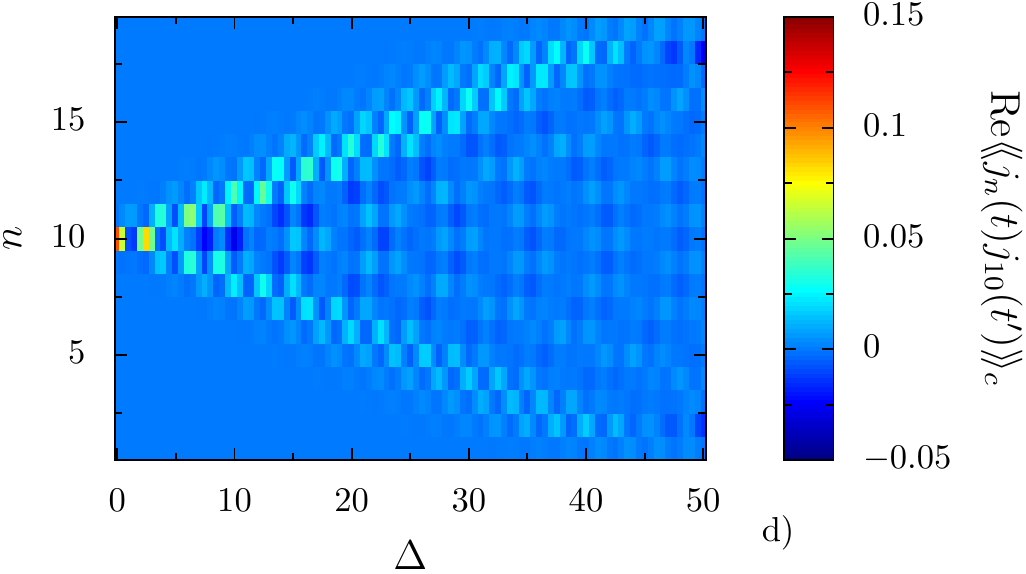}
\end{minipage}
\caption{Real part $\mathrm{Re}\llangle j_n(t)j_{10}(t')\rrangle_c$ of covariance  from the SLED approach for a chain with same parameters as used for figure~\ref{fig:j_over_eta}. The reservoirs' temperatures differ with $\kb T_1=2.0$ and $\kb T_N=0$, which breaks the model's symmetry. The density plots show results for different damping strengths  $\gamma=10^{-3}$ (a), $\gamma=0.3$ (b) and $\gamma=1.5$ (d) while all couplings are $\mu=0.3$. The effects from the nonequilibrium situation are most apparent when the couplings $\mu$ 'match' the damping $\gamma$. This is highlighted in plot c) showing  $\mathrm{Re}\llangle j_n(t)j_{10}(t')\rrangle_c$ over $n$ for $\Delta=20$ where the asymmetries of the covariance with respect to the reference site $n=10$ are largest for $\gamma=0.3$ (purple), while for $\gamma=10^{-3}$ (blue) and $\gamma=1.5$ (red) these effects are suppressed.}
\label{fig:noneq_SLED_comp_coup}
\end{figure*}

In figure~\ref{fig:noneq_SLED_comp_coup} we study nonequilibrium effects achieved by the coupling of a system with $N=20$ oscillators to two reservoirs with different temperatures $\kb T_1=2.0$ and $\kb T_N=0$. While plot a) shows the real part of the covariance  $\llangle j_n(t)j_{10}(t')\rrangle_c$ for very weak damping  $\gamma=10^{-3}$, panel b) shows the covariance achieved with  $\gamma=0.3$ and d) shows $\mathrm{Re}\llangle j_n(t)j_{10}(t')\rrangle_c$ for $\gamma = 1.5$. When comparing the covariances plotted over $n$, as exemplary done for $\Delta=20$ in c), one immediately sees that the impact of the temperature gradient applied to the system is most distinct for $\gamma = 0.3$. This is the damping $\gamma$ providing the best 'match' with the couplings within the chain $\mu$ and corresponds to the vicinity of the maximum of the steady-state heat flux $\llangle j\rrangle$ shown in figure~\ref{fig:j_over_eta}. Even the steady-state flux and the covariances of the heat fluxes are different observables, they both are sensitive with respect to the coupling strengths to the heat baths. This shows that a method with validity beyond the weak coupling regime is crucial in order to study nonequilibrium effects on heat-flux correlations.

The ability to study heat-flux correlations in nonequilibrium qualifies our method to be used for the quantification of transport in two respects. In order to decide if a system shows ballistic, diffusive or localized behavior one can consider a nonequilibrium situation and calculate the heat flux versus the chain length. This reveals an exponential or algebraic dependence and allows to extract localization length scales. Second, one can use second order correlations in thermal equilibrium as done for anharmonic classical systems in order to classify transport properties \cite{Zhao_2006, Liu_2014, Li_2015}. With the formalism presented here, we provide a generalization to the quantum regime and to situations out of equilibrium. According to \cite{Zhao_2006, Liu_2014, Li_2015} of specific relevance are momentum and energy correlations with momentum correlations being inherent to fluctuations in the heat-flux (see (\ref{eq:real_covsled}) and (\ref{eq:im_covsled})), while energy fluctuations can be derived in a similar manner as the correlations of heat we targeted here. The impact of out of equilibrium situations on transport parameters such as localization length and diffusion exponent obtained in equilibrium will be studied in future work. 

\section{Conclusions}

Although the damped harmonic oscillator has been studied extensively, a correct description of system reservoir interactions particularly for strong damping is challenging but essential when computing thermodynamic quantities such as heat fluxes. Based on the stochastic Liouville-von Neumann equation we presented an efficient computational scheme for damped systems which proved to be reliable over the full range of damping strengths and temperatures. We exploited this flexibility to study heat currents also for stronger coupling to thermal reservoirs and observed a similar behavior as found in approaches based on Langevin equations, namely, a distinct maximum for the currents when the reservoir coupling strength is tuned. Its position depends on the intra-chain couplings.

As a main result the scheme also allows us to gain spatiotemporal correlations of heat fluxes. For weak damping, excellent agreement is found with analytical calculations in a canonical Gibbs ensemble. However beyond this domain substantial deviations occur. The parameters for which we found the maximum in the currents led to the most distinct nonequilibrium effects in the fluctuations. The most distinct nonequilibrium effects in the correlations appear in ranges of parameter space where mean currents exhibit maxima. 

The present approach can now easily be extended to disordered systems. The computational efficiency allows one to study heat fluxes in large chains with randomization of any system parameter. The spatiotemporal correlation patterns might enable one to investigate localization and diffusion solely in time-domain which would represent an alternative to methods based on a transformation to normal modes. 
Moreover, an extension to driven systems where the external driving can either couple linearly or quadratically to the system's position is possible. The latter represents the important case of parametric driving and is of particular relevance for nano heat engines. For example, the spatial motion of an ion in a linear Paul trap was induced in \cite{Abah_2012, Rossnagel_2016} by a periodic narrowing and widening of the system's ground state frequency. In these situations, $\mathbf{M}$ from (\ref{eq:1st_cumulants}) becomes time dependent and the Green's function in (\ref{eq:greenfunc}) does not remain time translational invariant. This demands to decompose the Green's function into a forward and backward propagator $\mathbf{G}(t,t')=\mathbf{G}_\mathrm{f}(t,0)\mathbf{G}_\mathrm{b}(0,t')$ with the equations of motion given by $\dot{\mathbf{G}}_\mathrm{f}(t,0)=\mathbf{M}(t)\mathbf{G}_\mathrm{f}(t,0)$ and $\dot{\mathbf{G}}_\mathrm{b}(0,t') = -\mathbf{G}_\mathrm{b}(0,t')\mathbf{M}(t')$ for the forward and backward propagation respectively. This formalism was used to study entanglement generation through local driving in a bipartite system in analogy to \cite{Schmidt2013} and we consider it to open ways to study different settings for quantum heat engines beyond the limitations of weak couplings and adiabatic driving modes.

\ack

We thank Udo Seifert for valuable discussions. This work was supported by Deutsche Forschungsgemeinschaft through grant AN336/6-1.

\appendix
\section{Equations of motion for trace-average cumulants}
\label{app:cumulants}
Based on~(\ref{eq:Ham_qu}) and (\ref{eq:adjoint_master}), we calculate the equations of motion for the first and second cumulants. We assume a general a model, where each oscillator $n$ can be coupled to a reservoir leading to a nonzero damping $\gamma_n$. For the site indexed with $n$ we get ($a_n = m\omega_0^2 +\mu_n+\mu_{n-1}$):

\begin{eqnarray}
\difft \langle q_n\rangle_c &= \frac{\langle p_n\rangle_c}{m} \label{eq:cumulants_dyn_qn}\\
\difft\langle p_n\rangle_c &= +\mu_{n-1}\langle q_{n-1}\rangle_c-a_n\langle q_n\rangle_c\nonumber\\
&- \gamma_n\langle p_n\rangle_c+\mu_n\langle q_{n+1}\rangle_c +\xi_n(t) \label{eq:cumulants_dyn_pn}\;,
\end{eqnarray}
where $\gamma_n$ and $\xi_n(t)$ are only non-zero if $n=1,N$ for the system displayed in figure~\ref{fig:osci_chain}. The couplings are indexed since the topology of a chain where $\mu_n$ connects the site $n$ with its neighbor $n+1$ is guaranteed by the boundary condition $\mu_0=\mu_N=0$. For $N$ modes, we have a matrix equation $\difft\langle\vec{\sigma}\rangle_c =\mathbf{M}\langle\vec{\sigma}\rangle_c+\vec{\xi}(t)$ with $\mathbf{M}$ being a $2N\times 2N$ matrix

\begin{equation}
\mathbf{M} =
\left(\begin{array}{cccccc}
0 & 1/m & 0 &\dots  & \dots & 0\\
-a_n & -\gamma_1 & \mu_1  & \dots & \dots & 0\\
\vdots  & \ddots & \ddots  &\ddots   &   & \vdots  \\
\vdots  &  &\ddots  & \ddots  &\ddots   & \vdots  \\
0 &\dots  & 0 & 0 & 0 & 1/m\\
0 &\dots  & \mu_{N-1} & 0 & -a_n & -\gamma_N
\end{array}\right)
\label{eq:M1}
\end{equation}
where the coefficients of the $n$th phase-space variables form a $2\times2$ block structure and the couplings appear in the off-diagonals.

The 2nd cumulants are equivalently calculated to the first and read for the site with index $n$:

\begin{eqnarray}
\difft\langle q_n^ 2\rangle_c =& \frac{2}{m}\langle q_n p_n\rangle_c \label{eq:cumulants_dyn_qnsquare}\\
\difft\langle p_n^ 2\rangle_c =& -2\gamma_n\langle p^2_n\rangle_c - 2a_n\langle q_n p_n\rangle_c\nonumber\\
& + 2\mu_{n-1}\langle q_{n-1}p_n\rangle_c+2\mu_n\langle q_{n+1}p_n\rangle_c \label{eq:cumulants_dyn_pnsquare}\\
\difft\langle q_nq_l\rangle_c =& \frac{\langle q_lp_n\rangle_c}{m}+ \frac{\langle q_np_l\rangle_c}{m}\label{eq:cumulants_dyn_qnpl}\\
\difft\langle p_np_l\rangle_c =& -a_n\langle q_np_l\rangle_c-a_n\langle q_lp_n\rangle_c - (\gamma_n+\gamma_l)\langle p_np_l\rangle_c\nonumber\\
& + \mu_n\langle q_{n+1}p_l\rangle_c + \mu_{n-1}\langle q_{n-1}p_l\rangle_c\nonumber\\
&+ \mu_l\langle q_{l+1}p_n\rangle_c + \mu_{l-1}\langle q_{l-1}p_n\rangle_c\label{eq:cumulants_dyn_pnpl}\\
\difft\langle q_n p_l\rangle_c =& -a_n\langle q_nq_l\rangle_c + \frac{\langle p_np_l\rangle_c}{m} -\gamma_l
\langle q_n p_l\rangle_c\nonumber\\
&+\mu_{l-1}\langle q_{l-1}q_n\rangle_c+\mu_l\langle q_nq_{l+1}\rangle_c \label{eq:cumulants_dyn_qnpn}\;.
\end{eqnarray}
The only entries in $\mathbf{M}$ which do not represent unitary evolution are the non-zero diagonal entries $-\gamma_l$. This allows us to rewite the preceding equations in the compact form of~(\ref{eq:2nd_cumulants}),
\[
\difft \upperind{\BSig}{tr} = \mathbf{M}\upperind{\BSig}{tr}  + \upperind{\BSig}{tr} \mathbf{M}^{\dagger}\;.
\]

\section{Propagation of spatiotemporal correlation functions}
\label{app:Corrs}

We derive the dynamics of the elements in $\BSig_>(\Delta)$ by considering~(\ref{eq:adjcorr}) for all combinations of phase-space variables. The result is in close analogy to the equations of \ref{app:cumulants}. However, it is simpler than the propagation leading to the covariance matrix, since the adjoint Liouvillian $\mathcal{L}^\dagger$ is applied only to one of the operators in the product. Abbreviating
\begin{equation}
\llangle A(\Delta) B(0)\rrangle = \llangle A \mathcal{V}(\Delta,0) B \mathcal{V}(0,t_0)\rrangle
\end{equation}
and considering $A\in \{q_n,p_n\}$, we find from~(\ref{eq:adjcorr})
\begin{eqnarray}
&\fl\partial_\Delta \llangle q_n(\Delta)B(0)\rrangle = \frac{1}{m}\llangle p_n(\Delta)B(0)\rrangle\label{eq:qqcor}\\
&\fl\partial_\Delta \llangle p_n(\Delta)B(0)\rrangle = \mu_{n-1}\llangle q_{n-1}(\Delta)B(0)\rrangle
-a_n\llangle q_n(\Delta)B(0)\rrangle\nonumber\\
&\fl-\gamma_n\llangle p_n(\Delta)B(0)\rrangle+\mu_n\llangle q_{n+1}(\Delta)B(0)\rrangle+\langle\xi_n(\Delta)\langle B(0)\opind{\rangle}{tr}\opind{\rangle}{stoch}\label{eq:pqcor}\;,
\end{eqnarray}
The last term in~(\ref{eq:pqcor}) represents elements of $\mathbf{z}(\Delta)$. It requires further treatment in order to arrive at closed equations of motion. Observing that
\begin{equation}
\langle\xi_n(\Delta)\langle B(0)\opind{\rangle}{tr}\opind{\rangle}{stoch}
=
\langle\xi_n(0)\langle B(-\Delta)\opind{\rangle}{tr}\opind{\rangle}{stoch}
\end{equation}
and applying the adjoint~(\ref{eq:adjoint_master}) to $B(-\Delta)$ leads to~(\ref{eq:corrpropz}).

\section*{References}
\bibliography{literature_paper_without_url}

\end{document}